\documentclass[conference,9pt]{IEEEtran}

\usepackage{cite}
\usepackage{amsmath,amssymb,amsfonts}
\usepackage{algorithmic}
\usepackage{graphicx}
\usepackage{textcomp}
\usepackage{xcolor}
\usepackage{tabularx}
\usepackage{multirow}
\usepackage{booktabs}
\usepackage{makecell} 
\usepackage{amssymb}
\usepackage{float}
\usepackage{subcaption}
\usepackage{pdfpages}

\def\BibTeX{{\rm B\kern-.05em{\sc i\kern-.025em b}\kern-.08em
    T\kern-.1667em\lower.7ex\hbox{E}\kern-.125emX}}
\begin{document}

\title{CoLLAP: \textbf{Co}ntrastive \textbf{L}ong-form \textbf{L}anguage-\textbf{A}udio \textbf{P}retraining with Musical Temporal Structure Augmentation}

\author{
\IEEEauthorblockN{Junda Wu}
\IEEEauthorblockA{\textit{Computer Science and Engineering} \\
\textit{UC  San Diego}\\
La Jolla, USA \\
juw069@ucsd.edu } \\
\IEEEauthorblockN{Amit Namburi}
\IEEEauthorblockA{\textit{Computer Science and Engineering} \\
\textit{UC  San Diego} \\
La Jolla, USA \\
anamburi@ucsd.edu } 

\and
\IEEEauthorblockN{Warren Li }
\IEEEauthorblockA{\textit{Computer Science and Engineering} \\
\textit{UC  San Diego}\\
La Jolla, USA \\
wyl003@ucsd.edu } \\
\IEEEauthorblockN{Carol Chen}
\IEEEauthorblockA{\textit{Computer Science Department} \\
\textit{UC  Los Angeles}\\
Los Angeles, USA \\
carolchen12@ucla.edu} 

\and
\IEEEauthorblockN{Zachary Novack}
\IEEEauthorblockA{\textit{Computer Science and Engineering} \\
\textit{UC  San Diego}\\
La Jolla, USA \\
znovack@ucsd.edu }  \\
\IEEEauthorblockN{Julian McAuley}
\IEEEauthorblockA{\textit{Computer Science and Engineering} \\
\textit{UC  San Diego}\\
La Jolla, USA \\
jmcauley@ucsd.edu }

}

\maketitle

\begin{abstract}
Modeling temporal characteristics plays a significant role in the representation learning of audio waveform.
We propose \textbf{Co}ntrastive \textbf{L}ong-form \textbf{L}anguage-\textbf{A}udio \textbf{P}retraining (\textbf{CoLLAP}) to 
significantly extend the perception window for both the input audio (up to 5 minutes) and the language descriptions (exceeding 250 words),
while enabling contrastive learning across modalities and temporal dynamics.
Leveraging recent Music-LLMs to generate long-form music captions for full-length songs, augmented with musical temporal structures,
we collect 51.3K audio-text pairs derived from the large-scale AudioSet training dataset, where the average audio length reaches 288 seconds.
We propose a novel contrastive learning architecture that fuses language representations with structured audio representations 
by segmenting each song into clips and extracting their embeddings. 
With an attention mechanism, we capture multimodal temporal correlations, 
allowing the model to automatically weigh and enhance the final fusion score for improved contrastive alignment.
Finally, we develop two variants of the CoLLAP model with different types of backbone language models.
Through comprehensive experiments on multiple long-form music-text retrieval datasets, 
we demonstrate consistent performance improvement in retrieval accuracy compared with baselines.
We also show the pretrained CoLLAP models can be transferred to various music information retrieval tasks, with heterogeneous long-form multimodal contexts.

\end{abstract}

\section{Introduction} \label{sec:intro}
\begin{figure}[htp]
    \centering

    \begin{subfigure}[b]{\linewidth}
        \centering
        \includegraphics[page=2, width=\linewidth]{figs/collap-temp.pdf}
        \caption{
        Illustration of conventional \textbf{CLAP} model, whose inputs include short music captions (less than 50 words) and short audio clips (less than 30 seconds).
        CLAP only extracts 1-dimensional global textual and audio embeddings to calculate cosine similarity.
        }
        \label{fig:1-1}
    \end{subfigure}
    
    \vspace{1em}
    
    \begin{subfigure}[b]{\linewidth}
        \centering
        \includegraphics[page=1, width=\linewidth]{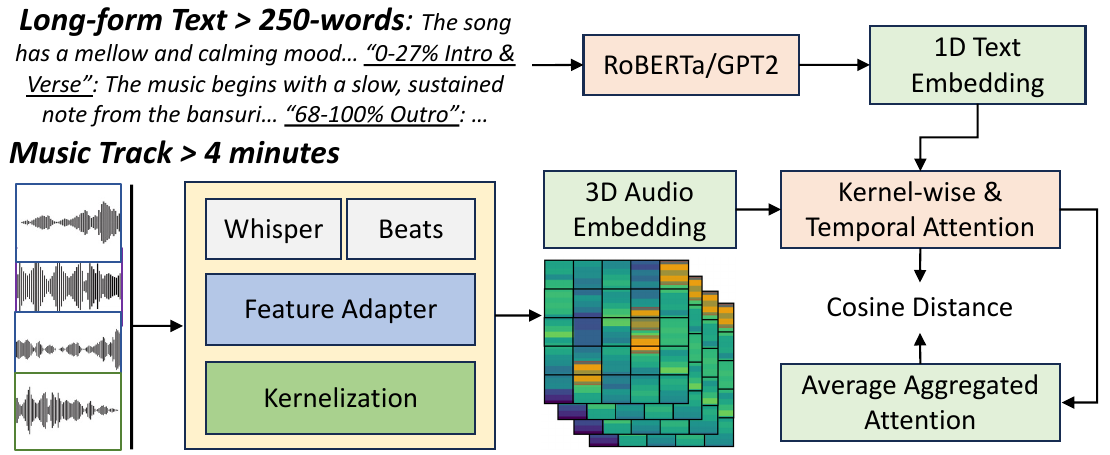}
        \caption{
        Illustration of our proposed \textbf{CoLLAP} model, whose inputs include fine-grained and temporally-aware music descriptions (more than 250 words) and full-length music tracks (more than 4 minutes).
        CoLLAP extracts 3-dimensional audio embeddings and aggregates using 3D-attention pooling that explicitly models temporal attention.
        We also enable two variants of CoLLAP using different language backbones RoBERTa and GPT2.}
        \label{fig:1-2}
    \end{subfigure}

    \caption{Comparison of conventional CLAP (Figure \ref{fig:1-1}) and our proposed CoLLAP (Figure \ref{fig:1-2}).}
    \label{fig:1}

    \vspace{-1em}
\end{figure}

The ability to effectively model temporal characteristics is essential in the representation learning of audio waveforms, 
especially for complex and full-length music tracks.
Music information retrieval works \cite{whiteley2006bayesian, weiss2011unsupervised} have studied approaches to extract musical temporal and structural information,
which can be further used to augment models' music understanding abilities \cite{wu2024futga}.
The recent contrastive learning approaches \cite{elizalde2023clap,zhu2024cacophony,wu2023large} enable to extract such information as latent audio representations,
which are trained to distinguish between matched text-audio pairs and other mismatched pairs by capturing distinctive features in the audio data (illustrated in Figure \ref{fig:1-1}).
However, such methods have focused on relatively short segments, limiting the model’s ability to handle longer, more nuanced sequences.

To address these challenges, we introduce Contrastive Long-form Language-audio Pretraining (CoLLAP), 
which extends the perception window to handle both long-form audio inputs and detailed language descriptions.
We illustrate the comparison between the conventional CLAP model and our proposed CoLLAP model in Figure \ref{fig:1}.
The CoLLAP model uses a feature extractor to segment music tracks into frames and encode each by a kernel function.
Then kernel-wise and temporal attention mechanisms are employed to measure global and temporal alignment between audio and text.
Finally, the model is optimized with contrastive learning using weighted similarity scores from both kernel-wise and temporal attention.
CoLLAP effectively extends the perception window for both the input audio (up to 5 minutes) and the language descriptions (exceeding 250 words),
which enables retrieval of full-length music tracks with fine-grained music descriptions.

To enable large-scale contrastive pretraining of CoLLAP, we leverage a Music-LLM augmented dataset of 51.3K audio-text pairs and 4,109 hours of audio,
derived from the large-scale AudioSet training data, with an average audio length of 288 seconds and an average text length of 256 words.
In addition, we develop two variants of CoLLAP based on two different backbone language models, 
Roberta-base \cite{liu2019roberta} and GPT2 \cite{radford2019language}.

Finally, we conduct comprehensive experiments on multiple long-form music-text retrieval datasets 
and observe consistent improvement in retrieval accuracy of CoLLAP compared with baseline models.
We also evaluate CoLLAP's transfer learning ability in various music information retrieval tasks that involve heterogeneous long-form multimodal contexts, 
including speech audio and Wikipedia free-form long-context.
In addition, we also observe better generalizability in the CoLLAP-GPT2 variant compared to RoBERTa model backbone 
due to the GPT2 model's better language modeling ability of long-context.
We summarize our contributions as follows:
\begin{itemize}
    \item We propose the Contrastive Long-form Language-audio Pretraining (CoLLAP) model for multimodal fusion and representation learning of long-form audio and language descriptions.
    \item We design a novel fusion mechanism that combines structured audio and language representations, leveraging attention to capture and weigh multimodal temporal correlations for improved contrastive alignment.
    \item We augment a dataset of 4,109 hours of long-form full-length music tracks, paired with musical structural augmented captions generated by Music-LLMs.
    \item Through comprehensive experiments we demonstrate that CoLLAP consistently outperforms baseline models in long-form text-audio retrieval, and show its generalizability across different tasks.
\end{itemize}

\section{CoLLAP: Model Design and Learning}\label{sec:model}
\begin{figure}[htp]
    \centering
    \includegraphics[width=.95\linewidth]{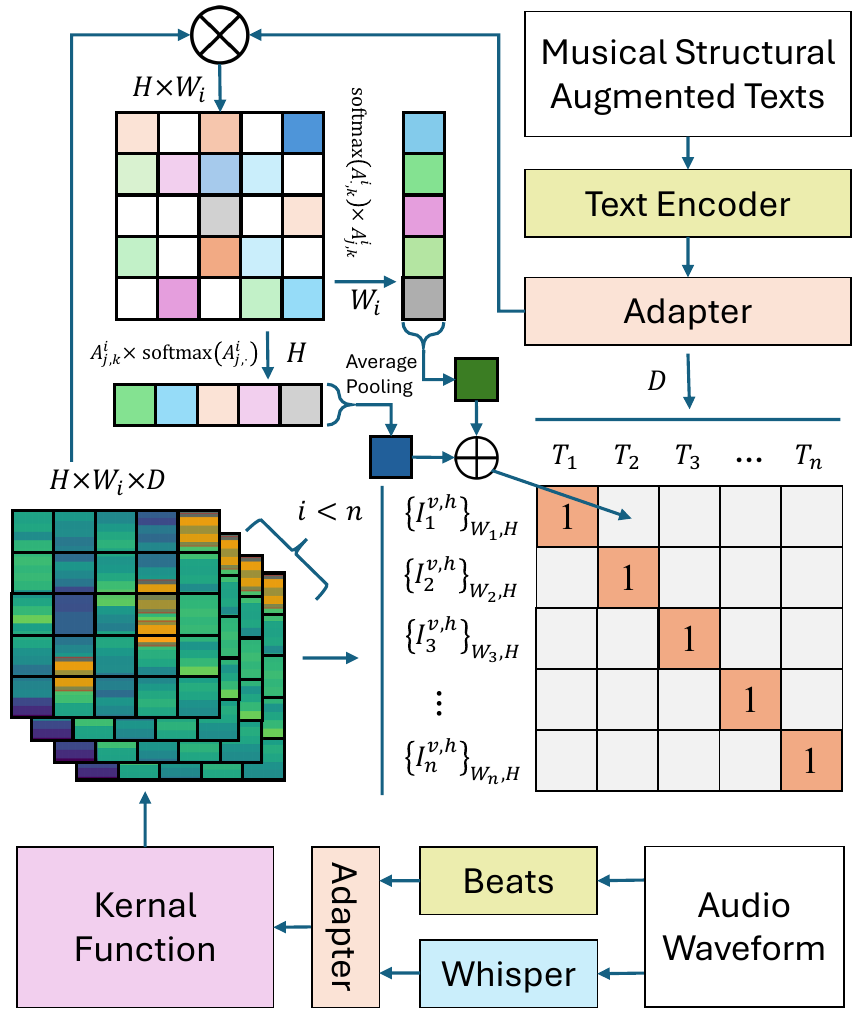}
    \caption{The model overview of CoLLAP. The input of backbone language models is musical structural augmented texts, 
    while audio waveform is encoded by the dual-feature extractor of Beats and Whisper models. 
    The encoded multimodal features are used for the calculation of temporal and kernel-wise attentions before computing contrastive learning loss.}
    \label{fig:model}
\end{figure}

We illustrate our CoLLAP model design in Figure~\ref{fig:model}, 
where full-length music track waveform is processed with a dual-feature extractor, 
while textual representations are extracted from musical structure augmented captions.
We split music tracks of variable lengths into frames to enable audio temporal attention with texts,
which extracts and measures both the global and temporal multimodal alignment scores.
With the temporal attention augmented alignment scores, 
we follow the conventional contrastive learning scheme \cite{radford2021learning,elizalde2023clap,yuan2024t,wu2023large},
where the contrastive loss will be propagated back to both the temporal attention and the feature extractors.

\subsection{Text and Dynamic Audio Encoders}
Given $N$ input audio-text pairs $\{(X_i, Y_i)\}_{i<N}$, 
we extract the textual embeddings $T_i\in \mathbb{R}^D$, musical embeddings $O_i\in \mathbb{R}^D$, and speech embeddings $S_i\in \mathbb{R}^D$ as follows:
\begin{equation*}
    T_i = f_T(Y_i;\theta_T), \quad
    O_i = f_O(X_i;\theta_O), \quad
    S_i = f_S(X_i;\theta_S),
\end{equation*}
where the model parameters of the text encoder $\theta_T$ are initialized from a pre-trained language model (\emph{e.g.}, RoBERTa \cite{liu2019roberta} and GPT-2 \cite{radford2019language}), while the music encoder and speech encoder are adapted from BEATS \cite{chen2022beats} and Whisper \cite{radford2023robust} models.
We fuse the musical and speech embeddings by an audio feature adapter linear layer $h_A$,
\begin{equation*}
    U_i = h_A\left([O_i,S_i]\right), I<N.
\end{equation*}
Then, we split the unified audio representation with a length of $T$ into consecutive frames with a kernel function with a kernel size of $H$ and stride step of $S_T$,
\begin{equation*}
    H = \left\lfloor \frac{T \cdot \eta_K}{30} \right\rfloor, \quad S_T = \left\lfloor \frac{T \cdot \eta_S}{30} \right\rfloor,
\end{equation*}
where $\eta_K$ is pre-defined to determine how many seconds per frame, and $\eta_S$ determines seconds per stride.
Finally, the processed audio representation is unfolded and reshaped to $I_i=\{I_i^{v,h}\}_{W_i,H}\in \mathbb{R}^{H\times W_i \times D}$,
\begin{equation}
     I_i = \text{Unfold}(U_i, H, S_T), \text{where}\; W_i = \left\lfloor \frac{T - H}{S_T} + 1 \right\rfloor.
\end{equation}
With the audio tokenized with fixed-length frames $I_i=\{I_i^{v,h}\}_{W_i,H}$, 
we can calculate kernel-wise attention and temporal attention to augment the multimodal alignment estimation.

\subsection{Multimodal and Temporal Attention Augmentation}
Given the audio representation $I_i=\{I_i^{v,h}\}_{W_i,H}$ and the text representation $T_j$,
we calculate their cosine similarity 
\begin{equation}
    M_{i,j}=\{(I_i^{v,h})^\top T_j\}_{W_i,H},
\end{equation}
in each frame $v<W_i$ and each kernel $h<H$.
To further measure the text's attention on the individual frame and kernel,
we calculate the kernel-wise attention $A_{i,j}^K$ and temporal attention $A_{i,j}^T$,
\begin{align}
    A_{i,j}^K(v,h) &= \frac{e^{M_{i,j}(v,h)}}{\sum_{k<H}e^{M_{i,j}(v,k)}}, \\
    A_{i,j}^T(v,h) &= \frac{e^{M_{i,j}(v,h)}}{\sum_{l<W_i}e^{M_{i,j}(l,h)}}, 
\end{align}
where $M_{i,j}(v,h)$ is the corresponding cosine similarity score of the $v$-th frame and $h$-th kernel in $M_{i,j}$.

\subsection{Temporal Attention Fused Contrastrive Learning}
Then we use the calculated kernel-wise attention $A_{i,j}^K$ and temporal attention $A_{i,j}^T$ to weigh and sum the original cosine similarity matrix $M_{i,j}$.
To obtain the global similarity between the text and audio, $M_{i,j}$ is weighted by the kernel-wise attention $A_{i,j}^K$ with an average pooling,
\begin{equation}
    r^K_{i,j} = \frac{1}{H}\sum_{k<H}\sum_{l<W_{i}} M_{i,j}(k,l)\cdot A_{i,j}^K(k,l).
\end{equation}
To capture the temporal attention-weighted similarity between text and audio, we further derive the similar similarity score,
\begin{equation}
    r^T_{i,j} = \frac{1}{W_i}\sum_{l<W_i}\sum_{k<H} M_{i,j}(k,l)\cdot A_{i,j}^T(k,l).
\end{equation}
Finally, we compose the two weighted similarity scores with two scalers $\gamma_K$ and $\gamma_T$ for balance. 
Therefore, each pairwise cosine similarity score $r_{i,j}\in \mathbb{R}^{N\times N}$ in the mini-batch is calculated as
\begin{equation}
    r_{i,j} = \gamma_K \cdot r^K_{i,j} + \gamma_T \cdot r^T_{i,j}.
\end{equation}
Following \cite{elizalde2023clap,radford2021learning,zhu2024cacophony}, we adopt the conventional contrastive loss function to derive the final loss,
\begin{equation}
    L = -\sum_{i<N}\log \frac{e^{r_{i,i}}}{\sum_{j<N} e^{r_{i,j}}},
\end{equation}
where the contrastive loss will be propagated back to both the temporal attention and the feature extractors.

\section{Long-form and Structural-aware Text-audio Retrieval Dataset}\label{sec:data}
We collect a large-scale long-form audio waveform dataset derived from the full-length tracks from the training subset of AudioSet \cite{gemmeke2017audio}.
We filter out audio tracks whose lengths are either less than 2 minutes or longer than 5 minutes,
accumulating to a total of 51.3K and $4,109.50$ hours of audio tracks with an average length of $288.25$ seconds per track.
To further pair the full-length audio tracks with long-form and fine-grained captions that comprehensively describe the entire track,
we leverage the FUTGA model \cite{wu2024futga} to generate dense captions, 
which provides both global caption and temporally-aware structural information.
Therefore, the generated dense captions have an average of $256.94$ words for each caption.

\begin{table}[htp]
\centering
\small
\caption{Comparison of the statistics of existing text-music retrieval datasets and CoLLAP.}
\label{tab:stats}
\begin{tabular}{lcccc}
\toprule
Dataset             & Pairs &  \makecell{Audio (hrs) \\ Duration } & \makecell{Ave. (secs)\\ Duration } & \makecell{Ave. \\ Words }  \\

\midrule
AudioCaps \cite{kim2019audiocaps}      & 51k  &  144.9 & 10.23 & 9.0 \\ 
MusicCaps \cite{agostinelli2023musiclm}      & 6k & 15.3 & 10.00 & 48.9\\ 
LAION-Audio \cite{wu2023large}    & 633.5k & 4325.39 & 24.58 & -- \\
LP-MusicCaps \cite{doh2023lp}   & 514k & 4283.10 & 30.00 & 37.3 \\ 
CoLLAP         & 51.3k & 4109.50 & 288.25 & 256.94 \\ 
\bottomrule

\end{tabular}
\end{table}

We compare our collected long-form and structural-aware text-audio retrieval dataset in Table~\ref{tab:stats},
where we show that our dataset has a comparable total length of existing large-scale text-audio datasets (\emph{e.g.}, LAION-Audio \cite{wu2023large} and LP-MusicCaps \cite{doh2023lp}).
In addition, we demonstrate that our audio lengths are about ten times longer than the existing dataset on average,
while our average text length is about five times longer than the fine-grained MusicCaps \cite{agostinelli2023musiclm}.

\section{Experiments}\label{sec:exp}

\begin{table*}[t]
\centering
\small
\caption{
The retrieval performance of three variants of Larger-CLAP and two variants of CoLLAP on four evaluation datasets. 
We report recall values of rank 5, 20, and 100 for text-to-music (T2M) and music-to-text (M2T) retrieval. 
The best values are highlighted in bold fonts, while the second-best values are underlined. 
}
\label{tab:main}
\begin{tabular}{cl|cc|cc|cc|cc|c}
\toprule
\multicolumn{2}{c|}{Dataset}        & \multicolumn{2}{c|}{SongDescriber} & \multicolumn{2}{c|}{MusicCaps}    & \multicolumn{2}{c|}{AudioSet-Eval} & \multicolumn{2}{c|}{HarmonixSet} & \multirow{2}{*}{Average}\\
    \cmidrule(lr){3-4}                      \cmidrule(lr){5-6}                   \cmidrule(lr){7-8}           \cmidrule(lr){9-10}   
Model                      & Metric & ~~T2A~~  & ~~A2T~~ & ~~T2A~~ & ~~A2T~~ & ~~T2A~~ & ~~A2T~~ & ~~T2A~~ & ~~A2T~~ &  \\ 
\midrule
\midrule
\multirow{3}{*}{\makecell{HSTAT\\(RoBERTa)}}      
                             & R@5  & 3.12 & 5.95 & 2.67 & 3.18 & 4.85 & 3.74 & 1.54 & 2.14 & 3.40 \\
                             & R@5  & 11.33 & 17.85 & 8.01 & 10.16 & 13.88 & 13.88 & 6.06 & 7.96 & 11.14 \\
                             & R@5  & 41.64 & 49.58 & 28.75 & 30.90 & 43.17 & 44.05 & 23.87 & 25.89 & 35.98 \\
\midrule 
\multirow{3}{*}{\makecell{Larger\\CLAP}} 
                            & R@5    &    6.65    &   9.77   &    3.11  &   4.97    &    5.73    &     8.37  &   10.48   &   9.05   &       7.27  \\
                           & R@20   &    16.86    &    25.92   &   9.86  &   15.27    &   15.86    &     21.81  &   25.24   &   28.10   &      19.87  \\
                           & R@100   &    52.97    &   63.88   &   31.50  &   42.70   &   48.68    &     55.29  &   75.71   &   76.67   &      55.93  \\ 
\midrule
\multirow{3}{*}{\makecell{Cacophony}} 
                            & R@5    &    5.92    &   5.14   &    2.15  &   2.67  &    4.20    &     2.91  &  2.38  &  0.48  &  3.23  \\
                           & R@20   &    16.75    &   18.73   &   5.93  &   6.15   &   8.88   &     9.36  &  8.79  &  2.14  &  9.59  \\
                           & R@100   &    48.70    &   62.64   & 19.35   &   24.69   &   35.97   &     38.84  &  34.20  &  10.21  &  34.33  \\ 
\midrule
\midrule
\multirow{3}{*}{\makecell{CoLLAP\\(RoBERTa)}} 
                           & R@5    &  50.28    &  40.50    &  15.19    &  9.54     &  72.68   &   75.55    & 21.37   & 19.35  & 38.05  \\
                           & R@20   &  75.92    &  70.25    &  36.65    &  20.53    &  91.18   &   91.85    & 40.73   & 39.66  & 58.34  \\
                           & R@100  &  96.60    &  93.34    &  69.50    &  43.73    &  98.89   &   99.11    & 74.10   & 71.25  & 80.81  \\ 
\midrule
\multirow{3}{*}{\makecell{CoLLAP\\(GPT-2)}}   
                           & R@5     &  49.15    &   42.91  &  17.35  &   10.26    &  76.87    &   79.51    &  20.42   &  18.76  &  39.40  \\
                           & R@20    &  77.19    &   69.12  &  36.96  &   21.35    &  92.95    &   93.61    &  41.33   &  40.49  &  58.12  \\
                           & R@100   &  97.16    &   93.20  &  69.50  &   44.86    &  99.77    &   99.77    &  76.12   &  73.75  &  81.76  \\ 
\bottomrule
\end{tabular}
\end{table*}

\subsection{Implementation Details}

We implement the CoLLAP model using PyTorch 2.2 framework, leveraging pre-trained RoBERTa and GPT-2 models for the text encoder and 
adapting BEATS and Whisper models for the music and speech encoders, respectively. 
We collect 51.3K long-form audio-text pairs derived from the original AudioSet-train dataset \cite{gemmeke2017audio}, 
with an average audio duration of 288 seconds and a text length of 257 words. 

We initialize RoBERTa or GPT-2 for the text encoder with pre-trained weights. 
The music and speech encoders are respectively adapted from BEATS and Whisper models and concatenated as the fused audio embedding.
The fused textual and audio embedding sizes are set to $512$.
We fine-tuned the full parameters of both the text encoder and audio embeddings,
using an AdamW optimizer with a learning rate of $1e-4$ and weight decay of $1e-5$. 
We use a batch size of $50$ and enable in-batch contrastive learning loss implemented by a cross-entropy loss function.
The contrastive learning process is set for 20 epochs, with a linear learning rate scheduling.
The training process leverages 2 NVIDIA A100 GPUs with 40GB of memory.

\subsection{Datasets and Baselines}
We evaluate the CoLLAP model on three text-audio retrieval tasks, where four datasets, 
SongDescriber \cite{manco2023song}, MusicCaps \cite{agostinelli2023musiclm}, AudioSet-Eval \cite{gemmeke2017audio}, and HarmonixSet \cite{nieto2019harmonix}, 
are used for general long-form text-to-audio retrieval.
To test the retrieval accuracy in the speech domain, we evaluate the VCTK dataset \cite{veaux2013voice} for long-context transcript to full speech retrieval.
Finally, we further evaluate the model's zero-shot generalizability in free-form music context collected from Wikipedia pages and enable wiki-to-music retrieval.

We compare our proposed CoLLAP model with three contrastive learning baselines for the main experiment in Table~\ref{tab:main}: 
\textbf{HSTAT (RoBERTa)} \cite{wu2023large} employs RoBERTa for textual encoding and incorporates the feature fusion mechanism and keyword-to-caption augmentation;
\textbf{Larger CLAP} \cite{wu2023large} further enhances the model performance on music and speech domains by expanded pre-training;
\textbf{Cacophony} \cite{zhu2024cacophony} enhances by a hierarchical attention mechanism and advanced fusion techniques to dynamically combine multi-scale features from both modalities.
For our method, we develop two model variants \textbf{CoLLAP (RoBERTa)} and \textbf{CoLLAP (GPT2)} using two different language model backbones.

\subsection{Long-form Text-audio Retrieval}

The long-form text-audio retrieval experiments are designed to evaluate the effectiveness of the CoLLAP model in aligning extended audio tracks with their corresponding textual descriptions. Retrieval performance is measured using recall at ranks 5, 20, and 100 for both text-to-audio (T2A) and audio-to-text (A2T) retrieval tasks.

As presented in Table~\ref{tab:main}, the CoLLAP variants outperform the baseline models across all datasets, particularly on SongDescriber and HarmonixSet. The attention mechanisms in CoLLAP enable the model to effectively capture the temporal and multimodal correlations, leading to significant improvements in retrieval accuracy. The RoBERTa-based CoLLAP variant demonstrates slightly higher performance, especially in A2T retrieval tasks.

\subsection{Zero-shot Transcript-speech Retrieval}

We also evaluate CoLLAP's zero-shot transfer performance on transcript-speech retrieval tasks using the VCTK dataset. 
This experiment assesses the model's capability to align spoken content with corresponding transcripts without additional fine-tuning. 
Table~\ref{tab:mir} reports retrieval performance for both T2A and A2T tasks at various recall ranks.

The results indicate that the CoLLAP model variants maintain robust retrieval accuracy in this zero-shot setting. 
The GPT-2 based variant outperforms the RoBERTa-based variant, suggesting that GPT-2's generative capabilities may better handle the variability in spoken language. 
These findings highlight CoLLAP's potential for applications in speech recognition and audio-text alignment.

\begin{table}[H]
\centering
\small
\caption{Speech and audio retrieval on the VCTK dataset \cite{veaux2016superseded}. We report Recall@$k$ metrics for text-to-audio (T2A) and audio-to-text (A2T) retrieval.}
\label{tab:mir}
\begin{tabular}{cl|cccc}
\toprule
\multicolumn{2}{c|}{Model}        & 
\multirow{2}{*}{\makecell{HSTAT\\(RoBERTa)}} & 
\multirow{2}{*}{\makecell{Larger\\CLAP}} & 
\multirow{2}{*}{\makecell{CoLLAP\\(RoBERTa)}} & 
\multirow{2}{*}{\makecell{CoLLAP\\(GPT2)}}  \\

Dataset              & Metric    \\ 
\midrule
\multirow{3}{*}{\makecell{VCTK\\(T2A)}} 
                        & R@5    & 0.87  & 1.22  & 0.87  & 1.40\\
                        & R@20   & 4.03  & 4.38  & 3.50  & 5.96   \\
                        & R@100  & 18.94 & 19.47 & 15.78 & 21.75 \\
\midrule
\multirow{3}{*}{\makecell{VCTK\\(A2T)}}      
                        & R@5    & 0.87  & 1.40  & 0.70  & 1.75\\
                        & R@20   & 3.33  & 5.43  & 3.15  & 5.61 \\
                        & R@100  & 18.59 & 18.42 & 16.31 & 21.05\\
\bottomrule
\end{tabular}
\end{table}

\subsection{Zero-shot Wiki-music Retrieval}

Finally, we assess CoLLAP's generalizability in retrieving music-related content from textual descriptions in a zero-shot manner using the Wiki-music dataset. 
This dataset includes Wikipedia articles paired with audio clips, and the task involves retrieving the correct audio clip given a text query and vice versa. 
The retrieval performance is detailed in Table~\ref{tab:wiki}.

CoLLAP achieves significant gains over the baseline models in the Wiki-SD and Wiki-MC tasks. 
The model's attention mechanisms allow it to effectively align long-form text with corresponding audio segments, leading to improved retrieval accuracy. 
These results suggest that CoLLAP can be effectively transferred to diverse music-related information retrieval tasks, 
making it a versatile tool for exploring large-scale multimodal datasets.

\begin{table}[H]
\centering
\small
\caption{Wikipedia context and audio retrieval on the MusicCaps and SongDescriber datasets. We report Recall@$k$ metrics for wiki-to-music (W2M) and music-to-wiki (M2W) retrieval.}
\label{tab:wiki}
\begin{tabular}{cl|cc|cc|c}
\toprule
\multicolumn{2}{c|}{Dataset}        & \multicolumn{2}{c|}{Wiki-SD} & \multicolumn{2}{c|}{Wiki-MC}  & \multirow{2}{*}{Average}\\
    \cmidrule(lr){3-4}                      \cmidrule(lr){5-6}                    
Model                      & Metric & W2M  & M2W & W2M & M2W &  \\ 
\midrule
\midrule
\multirow{3}{*}{\makecell{HSTAT\\(RoBERTa)}} 
                            & R@5    &  3.12 & 4.67 & 3.90 & 3.80 & 3.87 \\
                            & R@20   &  9.92 & 14.59 & 10.47 & 10.68 & 11.41 \\
                            & R@100  & 37.82 & 43.63 & 31.42 & 27.93 & 35.20 \\
\midrule
\multirow{3}{*}{\makecell{Larger\\CLAP}}      
                             & R@5     & 4.24   & 8.92  & 4.10  & 6.05 & 5.83  \\
                             & R@20    & 14.73  & 25.21 &  13.24  & 17.14 & 17.58 \\
                             & R@100   & 45.60  & 55.52 & 39.73  & 43.32 & 46.05 \\ 
\midrule
\midrule
\multirow{3}{*}{\makecell{CoLLAP\\(RoBERTa)}}
                           & R@5    &  39.51 &  34.70 & 9.03  & 7.08 & 22.59  \\
                           & R@20   &  60.33 &  59.77 & 24.33 & 14.78 & 39.81 \\
                           & R@100  &  79.74 &  75.21 & 49.58 & 35.31 & 59.97 \\ 
\midrule
\multirow{3}{*}{\makecell{CoLLAP\\(GPT-2)}}   
                           & R@5     & 39.37 & 36.96 & 9.75  & 7.90  & 23.50 \\
                           & R@20    & 61.18 & 57.93 & 22.68 & 14.57 & 39.10 \\
                           & R@100   & 80.45 & 74.64 & 46.61 & 33.05 & 58.69 \\ 
\bottomrule
\end{tabular}
\end{table}

\section{Conclusion}\label{sec:conclusion}
In this paper, we introduce CoLLAP, a novel contrastive learning framework designed for long-form language-audio representation learning. Our model leverages dual-feature extraction and a multimodal attention mechanism to effectively capture both global and temporal alignments between lengthy audio tracks and detailed textual descriptions. Through comprehensive experiments across multiple datasets, including SongDescriber, MusicCaps, AudioSet-Eval, HarmonixSet, and Wiki-music, we demonstrate that CoLLAP significantly improves retrieval performance over existing baseline models.

\bibliographystyle{IEEEtran}
\bibliography{IEEEabrv,IEEEfull}

\begin{thebibliography}{10}
\providecommand{\url}[1]{#1}
\csname url@samestyle\endcsname
\providecommand{\newblock}{\relax}
\providecommand{\bibinfo}[2]{#2}
\providecommand{\BIBentrySTDinterwordspacing}{\spaceskip=0pt\relax}
\providecommand{\BIBentryALTinterwordstretchfactor}{4}
\providecommand{\BIBentryALTinterwordspacing}{\spaceskip=\fontdimen2\font plus
\BIBentryALTinterwordstretchfactor\fontdimen3\font minus
  \fontdimen4\font\relax}
\providecommand{\BIBforeignlanguage}[2]{{%
\expandafter\ifx\csname l@#1\endcsname\relax
\typeout{** WARNING: IEEEtran.bst: No hyphenation pattern has been}%
\typeout{** loaded for the language `#1'. Using the pattern for}%
\typeout{** the default language instead.}%
\else
\language=\csname l@#1\endcsname
\fi
#2}}
\providecommand{\BIBdecl}{\relax}
\BIBdecl

\bibitem{whiteley2006bayesian}
N.~Whiteley, A.~T. Cemgil, and S.~J. Godsill, ``Bayesian modelling of temporal
  structure in musical audio.'' in \emph{ISMIR}, 2006, pp. 29--34.

\bibitem{weiss2011unsupervised}
R.~J. Weiss and J.~P. Bello, ``Unsupervised discovery of temporal structure in
  music,'' \emph{IEEE Journal of Selected Topics in Signal Processing}, vol.~5,
  no.~6, pp. 1240--1251, 2011.

\bibitem{wu2024futga}
J.~Wu, Z.~Novack, A.~Namburi, J.~Dai, H.-W. Dong, Z.~Xie, C.~Chen, and
  J.~McAuley, ``Futga: Towards fine-grained music understanding through
  temporally-enhanced generative augmentation,'' \emph{arXiv preprint
  arXiv:2407.20445}, 2024.

\bibitem{elizalde2023clap}
B.~Elizalde, S.~Deshmukh, M.~Al~Ismail, and H.~Wang, ``Clap learning audio
  concepts from natural language supervision,'' in \emph{ICASSP 2023-2023 IEEE
  International Conference on Acoustics, Speech and Signal Processing
  (ICASSP)}.\hskip 1em plus 0.5em minus 0.4em\relax IEEE, 2023, pp. 1--5.

\bibitem{zhu2024cacophony}
G.~Zhu and Z.~Duan, ``Cacophony: An improved contrastive audio-text model,''
  \emph{arXiv preprint arXiv:2402.06986}, 2024.

\bibitem{wu2023large}
Y.~Wu, K.~Chen, T.~Zhang, Y.~Hui, T.~Berg-Kirkpatrick, and S.~Dubnov,
  ``Large-scale contrastive language-audio pretraining with feature fusion and
  keyword-to-caption augmentation,'' in \emph{ICASSP 2023-2023 IEEE
  International Conference on Acoustics, Speech and Signal Processing
  (ICASSP)}.\hskip 1em plus 0.5em minus 0.4em\relax IEEE, 2023, pp. 1--5.

\bibitem{liu2019roberta}
Y.~Liu, ``Roberta: A robustly optimized bert pretraining approach,''
  \emph{arXiv preprint arXiv:1907.11692}, 2019.

\bibitem{radford2019language}
A.~Radford, J.~Wu, R.~Child, D.~Luan, D.~Amodei, I.~Sutskever \emph{et~al.},
  ``Language models are unsupervised multitask learners,'' \emph{OpenAI blog},
  vol.~1, no.~8, p.~9, 2019.

\bibitem{radford2021learning}
A.~Radford, J.~W. Kim, C.~Hallacy, A.~Ramesh, G.~Goh, S.~Agarwal, G.~Sastry,
  A.~Askell, P.~Mishkin, J.~Clark \emph{et~al.}, ``Learning transferable visual
  models from natural language supervision,'' in \emph{International conference
  on machine learning}.\hskip 1em plus 0.5em minus 0.4em\relax PMLR, 2021, pp.
  8748--8763.

\bibitem{yuan2024t}
Y.~Yuan, Z.~Chen, X.~Liu, H.~Liu, X.~Xu, D.~Jia, Y.~Chen, M.~D. Plumbley, and
  W.~Wang, ``T-clap: Temporal-enhanced contrastive language-audio
  pretraining,'' \emph{arXiv preprint arXiv:2404.17806}, 2024.

\bibitem{chen2022beats}
S.~Chen, Y.~Wu, C.~Wang, S.~Liu, D.~Tompkins, Z.~Chen, and F.~Wei, ``Beats:
  Audio pre-training with acoustic tokenizers,'' \emph{arXiv preprint
  arXiv:2212.09058}, 2022.

\bibitem{radford2023robust}
A.~Radford, J.~W. Kim, T.~Xu, G.~Brockman, C.~McLeavey, and I.~Sutskever,
  ``Robust speech recognition via large-scale weak supervision,'' in
  \emph{International conference on machine learning}.\hskip 1em plus 0.5em
  minus 0.4em\relax PMLR, 2023, pp. 28\,492--28\,518.

\bibitem{gemmeke2017audio}
J.~F. Gemmeke, D.~P. Ellis, D.~Freedman, A.~Jansen, W.~Lawrence, R.~C. Moore,
  M.~Plakal, and M.~Ritter, ``Audio set: An ontology and human-labeled dataset
  for audio events,'' in \emph{2017 IEEE international conference on acoustics,
  speech and signal processing (ICASSP)}.\hskip 1em plus 0.5em minus
  0.4em\relax IEEE, 2017, pp. 776--780.

\bibitem{kim2019audiocaps}
C.~D. Kim, B.~Kim, H.~Lee, and G.~Kim, ``Audiocaps: Generating captions for
  audios in the wild,'' in \emph{Proceedings of the 2019 Conference of the
  North American Chapter of the Association for Computational Linguistics:
  Human Language Technologies, Volume 1 (Long and Short Papers)}, 2019, pp.
  119--132.

\bibitem{agostinelli2023musiclm}
A.~Agostinelli, T.~I. Denk, Z.~Borsos, J.~Engel, M.~Verzetti, A.~Caillon,
  Q.~Huang, A.~Jansen, A.~Roberts, M.~Tagliasacchi \emph{et~al.}, ``Musiclm:
  Generating music from text,'' \emph{arXiv preprint arXiv:2301.11325}, 2023.

\bibitem{doh2023lp}
S.~Doh, K.~Choi, J.~Lee, and J.~Nam, ``Lp-musiccaps: Llm-based pseudo music
  captioning,'' \emph{arXiv preprint arXiv:2307.16372}, 2023.

\bibitem{manco2023song}
I.~Manco, B.~Weck, S.~Doh, M.~Won, Y.~Zhang, D.~Bodganov, Y.~Wu, K.~Chen,
  P.~Tovstogan, E.~Benetos \emph{et~al.}, ``The song describer dataset: a
  corpus of audio captions for music-and-language evaluation,'' \emph{arXiv
  preprint arXiv:2311.10057}, 2023.

\bibitem{nieto2019harmonix}
O.~Nieto, M.~C. McCallum, M.~E. Davies, A.~Robertson, A.~M. Stark, and
  E.~Egozy, ``The harmonix set: Beats, downbeats, and functional segment
  annotations of western popular music.'' in \emph{ISMIR}, 2019, pp. 565--572.

\bibitem{veaux2013voice}
C.~Veaux, J.~Yamagishi, and S.~King, ``The voice bank corpus: Design,
  collection and data analysis of a large regional accent speech database,'' in
  \emph{2013 international conference oriental COCOSDA held jointly with 2013
  conference on Asian spoken language research and evaluation
  (O-COCOSDA/CASLRE)}.\hskip 1em plus 0.5em minus 0.4em\relax IEEE, 2013, pp.
  1--4.

\bibitem{veaux2016superseded}
C.~Veaux, J.~Yamagishi, K.~MacDonald \emph{et~al.}, ``Superseded-cstr vctk
  corpus: English multi-speaker corpus for cstr voice cloning toolkit,'' 2016.

\end{thebibliography}

\end{document}